\def\dAlem{{\vcenter{\hrule height.5pt
		\hbox{\vrule width.5pt height8pt \kern8pt
			\vrule width.5pt}
		\hrule height.5pt}}} 
\def\matth{\mathsurround=0pt }
\def\leftrightarrowfill{$\matth \mathord\leftarrow \mkern-6mu
  \cleaders\hbox{$\mkern-2mu \mathord- \mkern-2mu$}\hfill
  \mkern-6mu \mathord\rightarrow$}
\def\overleftrightarrow#1{\vbox{\ialign{##\crcr
     \leftrightarrowfill\crcr\noalign{\kern-1pt\nointerlineskip}
     $\hfil\displaystyle{#1}\hfil$\crcr}}}
\def\half{{\textstyle{1\over2}}}
\def\fourth{{\textstyle{1\over4}}}
\def\qentry#1{{q\over A-(#1+\nu)^2}}
\def\zentry{{q\over A-\nu^2}}
\parindent=1cm 
\raggedbottom
\centerline{\bf ``Natural'' Vacua in Hyperbolic Friedmann-Robertson-Walker
Spacetimes}
\centerline{Ian H.~Redmount\footnote{*}{E-mail:  redmount@hypatia.slu.edu}}
\centerline{\sl Department of Physics}
\centerline{\sl Parks College of Engineering and Aviation}
\centerline{\sl Saint Louis University}
\centerline{\sl PO Box 56907}
\centerline{\sl St.~Louis, Missouri~~63156--0907}
\vskip\baselineskip
\centerline{\bf ABSTRACT}\nobreak
Recent evidence indicates that the Universe is open, i.e., spatially
hyperbolic, longstanding theoretical preferences to the contrary
notwithstanding.  This makes it possible to select a vacuum state,
Fock space, and particle definition for a quantized field, by requiring
concordance with ordinary flat-spacetime theory at late times.  The
particle-number basis states thus identified span the physical state space
of the field at all times.  This construction is demonstrated here explicitly
for a massive, minimally coupled, linear scalar field in an open,
radiation-dominated Friedmann-Robertson-Walker spacetime.
\vskip\baselineskip
\noindent PACS number(s):  04.62.+v, 98.80.-k
\vfil\eject
\parskip=14pt
\centerline{\bf I.~~INTRODUCTION}\par\nobreak
For more than twenty years it has been well known that the definitions
of a vacuum state and ``particles'' for a quantized field in curved
spacetime are problematical.  Not that a Fock-basis space of particle-number
eigenstates cannot be defined for such a field; rather these can be defined
in an infinity of inequivalent ways.  It is not even necessary to consider
gravitation to find this problem, which arises even in flat spacetime.

For a similar length of time it had been widely held that the Universe,
as approximated by a homogeneous, isotropic Friedmann-Robertson-Walker
spacetime geometry, must be spatially flat.  (It was not always so; for
many years a model with closed, positively curved spatial sections was
believed to be the only natural choice~[1].)  The preference for a spatially
flat model was upheld either as a consequence of inflationary cosmology
or a motivation for it, or on the basis of various considerations of
``naturalness.''  Spatially flat or positive-curvature models require
a cosmological mass density at or above critical density; this underlies
the enormous body of theoretical and observational work on the cosmological
dark-matter problem.  But the weight of evidence, greatly strengthened by
recent observations, indicates that the Universe is not spatially flat, nor
positively curved---it is best described by an open model with hyperbolic
spatial sections, theoretical preferences to the contrary notwithstanding~[2,3].

Remarkably, these two conundra may be related.  It proves easy to select
a particularly simple vacuum state for a quantized field in a hyperbolic
Friedman-Robertson-Walker (FRW) spacetime; the corresponding particle-number
eigenstates span the state space of the field.  Conversely, hyperbolic (open)
FRW models are distinguished from spatially flat and positive-curvature models
in admitting such a ``natural'' definition of particles.

The choice of this ``natural'' vacuum state arises from the asymptotic behavior
of the hyperbolic FRW spacetime.  In the absence of a cosmological constant,
any such spacetime approaches the empty Milne spacetime geometry~[4,5] at late
times, as the expansion of the model drives its mass/energy density to zero.
But Milne spacetime is a portion of flat (Minkowski) spacetime.  A vacuum
choice can be made for a quantized field in Milne spacetime which is
equivalent to the familiar vacuum state of flat-spacetime theory, with
all of its desirable physical properties~[6].  Hence, a vacuum state is
selected on the original spacetime:  that which approaches this ``physical''
Milne vacuum at late times, thus inheriting all of its features.  The choice
of a vacuum determines a Fock space of states for the field; basis states
for this space define particles in the spacetime.

Of course this defines an ``out'' vacuum and ``out'' particles, notions
long familiar in curved-spacetime quantum field theory.  But in an open
Universe (with ordinary matter/energy content), the asymptotically Minkowskian
``out'' region is physical.  The behavior of a quantized field {\it must\/}
accord with familiar Minkowski-spacetime field theory there.  The state of
the field must lie in the Fock space spanned by the usual Minkowski vacuum
and many-particle states in the late-time limit, hence, by this ``natural''
choice of basis states.  Since both the actual field state and the basis
states are exact solutions of the field theory, the actual state must be
the same superposition of basis states at all times.  That is, this choice
of basis states spans the physical state space of the field, at all times.

This basis-state construction is illustrated here by means of an example:
a massive, minimally coupled, linear scalar field in radiation-dominated
FRW spacetime.  (The present Universe is matter-dominated, but this example
is simpler.)  The relevant spacetime geometries and quantum field theory
are described in Sec.~II.  The choice of a vacuum state for the field in
Milne spacetime, equivalent to the familiar Minkowski-spacetime vacuum,
is shown in Sec.~III.  The selection of a vacuum state for the field in
the radiation-dominated spacetime, which matches this Milne vacuum at late
times, is demonstrated in Sec.~IV.  The results are discussed in Sec.~V.
I use units here with $\hbar=c=1$, and sign conventions and general notation
as in Misner, Thorne, and Wheeler~[7].  
\vskip\baselineskip
\centerline{\bf II.~~FUNDAMENTALS}\par\nobreak
\centerline{\bf A.~~Spacetime geometries}\par\nobreak
The spacetimes of interest here are open Friedman-Robertson-Walker geometries,
with hyperbolic spatial sections and curvature parameter~$k=-1$.  They have
metrics of the form
$$ds^2=-dt^2+a^2(t)[d\chi^2+\sinh^2\chi(d\theta^2+\sin^2\theta\,d\phi^2)]\ ,
\eqno(2.1)$$
with $t\in[0,+\infty)$, $\chi\in[0,+\infty)$, $\theta\in[0,\pi]$, and
$\phi\in[0,2\pi)$.  The scale factor $a(t)$ completely characterizes each
spacetime.

The Milne universe~[4,5] is identified by the scale factor
$$a_M(t)=t\ .\eqno(2.2)$$
The resulting geometry corresponds, via the Einstein field equations,
to an empty Universe.  It is simply a portion of Minkowski spacetime, in
coordinates tied to observers moving radially with all possible velocities.
In coordinates $T=t\,\cosh\chi$, $R=t\,\sinh\chi$, $\theta$, and $\phi$,
metric~(2.1) with scale factor~$a_M$ takes on the familiar flat-spacetime
form.

The radiation-dominated FRW model has a scale factor given parametrically by
$$\eqalignno{a_R(\eta)&=a_*\,\sinh\eta&(2.3{\rm a})\cr
		t(\eta)&=a_*(\cosh\eta-1)\ ,&(2.3{\rm b})\cr}$$
with conformal-time parameter $\eta\in[0,+\infty)$, and $a_*$ a constant.
According to the Einstein field equations, the resulting geometry corresponds
to a density which decreases with time as~$a_R^{-4}$ and a pressure one-third
of this density; hence ``radiation dominated.''  For $\eta\gg1$ the density
and pressure rapidly approach zero, and $a_R$ approaches the Milne form~$a_M$.
\vskip\baselineskip
\centerline{\bf B.~~A quantized scalar field}\par\nobreak
A suitable example is a real scalar field~$\varphi$ interacting only with the
spacetime geometry, with mass~$\mu$ and minimal coupling.  The corresponding
field equation is
$$(\dAlem-\mu^2)\,\varphi=0\ ,\eqno(2.4)$$
with $\dAlem$ the covariant d'Alembertian corresponding to metric~(2.1).
The quantized field is expanded in terms of normal-mode solutions of this
equation and operator coefficients:
$$\eqalignno{\varphi&=\sum_{l,m}\int_0^\infty d\kappa\left\{b_{\kappa lm}\,
u_\kappa(t)\,\left[\prod_{j=0}^l(j^2+\kappa^2)^{1/2}\right](\sinh\chi)^{-1/2}
P_{-1/2+i\kappa}^{-(l+1/2)}(\cosh\chi)\,Y_{lm}(\theta,\phi)
+{\rm H.c.}\right\}\ .&\cr
&&(2.5)\cr}$$
Here $Y_{lm}$ is a spherical harmonic, and $P_{1/2+i\kappa}^{-(l+1/2)}$
is an associated Legendre function; the parameter~$\kappa$ represents
radial momentum in the hyperbolic spatial sections of the spacetime.
The time dependence $u_\kappa(t)$ is a solution of the equation
$${d^2u_\kappa\over dt^2}+{3\over a}{da\over dt}{du_\kappa\over dt}
+\left(\mu^2+{1+\kappa^2\over a^2}\right)\,u_\kappa=0\eqno(2.6)$$
satisfying the ``positive frequency'' normalization condition
$$ia^3\left(u_\kappa^*\overleftrightarrow{\partial\over\partial t}u_\kappa
\right)=1\ .\eqno(2.7)$$
Consequently the canonical commutation relations for the field~$\varphi$
impose on the operators~$b_{\kappa lm}$ and their Hermitian conjugates
commutation relations appropriate to annihilation and creation operators,
respectively.  The vacuum state of the field is defined by the condition
$b_{\kappa lm}|0\rangle=0$ for all modes~$\kappa lm$; a Fock-space basis
of particle states is obtained by applying the creation
operators~$b_{\kappa lm}^\dagger$ to this vacuum state.

The positive-frequency functions~$u_\kappa$ are not uniquely determined
by Eqs.~(2.6) and~(2.7).  In general there is a one-complex-parameter
family of such functions for each mode.  Each choice of~$u_\kappa$
corresponds to a different operator~$b_{\kappa lm}$, which defines
a different vacuum and particle states.  The Fock spaces spanned by
these various bases can be unitarily inequivalent.
\vskip\baselineskip
\centerline{\bf III.~~THE ``PHYSICAL'' VACUUM IN MILNE SPACETIME}\par\nobreak
For the Milne geometry, Eq.~(2.6) assumes a familiar form; a general solution
can be written
$$u_\kappa(t)={1\over t}\left[c_\kappa^{(1)}\,H_{i\kappa}^{(1)}(\mu t)
+c_\kappa^{(2)}\,H_{i\kappa}^{(2)}(\mu t)\right]\ ,\eqno(3.1)$$
where $c_\kappa^{(1,2)}$ are constants and $H_{i\kappa}^{(1,2)}$ are
Hankel functions.  Condition~(2.7) imposes the constraint
$${4\over\pi}\left(e^{-\pi\kappa}|c_\kappa^{(2)}|^2
-e^{\pi\kappa}|c_\kappa^{(1)}|^2\right)=1\eqno(3.2)$$
on the constants.  Each choice of $c_\kappa^{(1)}$, then, identifies a
different positive-frequency mode function, giving rise to a different
definition of vacuum and particle states.

There is a unique choice of coefficients~$c_\kappa^{(1,2)}$, however, for
which the resulting mode functions consist of superpositions entirely of
positive-frequency Minkowski-spacetime mode functions.  That choice is
$$c_\kappa^{(1)}=0\qquad\hbox{and}\qquad c_\kappa^{(2)}={\sqrt{\pi}\over2}\,
e^{-\pi\kappa/2}\ .\eqno(3.3)$$
A standard integral representation of the Hankel functions can be used to
show
$${1\over t}\,H_{i\kappa}^{(2)}(\mu t)\,(\sinh\chi)^{-1/2}\,
P_{-1/2+i\kappa}^{-(l+1/2)}(\cosh\chi)=\int_0^\infty\alpha_{\kappa l}(\beta)
\,e^{-i\omega T}\,j_l(kR)\,dk\ ,\eqno(3.4{\rm a})$$
with $k=\mu\sinh\beta$, $\omega=\mu\cosh\beta$,
$$\openup1\jot
\eqalignno{\alpha_{\kappa l}(\beta)&={-i\over\kappa^2+l^2}\left[
{\partial\over\partial\beta}+\tanh\beta-(l+1)\coth\beta\right]
\alpha_{\kappa,l-1}(\beta)\ ,&(3.4{\rm b})\cr
\hbox{and}\qquad\alpha_{\kappa0}(\beta)&=-\left({2\over\pi}\right)^{1/2}
{e^{-\pi\kappa/2}\over\pi\kappa/2}\sin(\kappa\beta)\tanh\beta\ .
&(3.4{\rm c})\cr}$$
In Eq.~(3.4a) $j_l$ is a spherical Bessel function; the integrand is the
time and radial dependence of a positive-frequency normal-mode solution of
Eq.~(2.4) in Minkowski coordinates $(T,R,\theta,\phi)$, with angular
momentum~$l$.  Consequently the Bogoliubov transformation between the
Milne-spacetime field decomposition with mode choice (3.3) and the usual
Minkowski-spacetime analysis is ``trivial,'' in the sense that it does
not mix positive- and negative-frequency mode functions or creation and
annihilation operators.  The vacuum state associated with mode choice~(3.3)
is the same state as the usual Minkowski-spacetime vacuum, with all its
symmetry, regularity, and minimal-energy properties.  Milne-spacetime
particle states built on this vacuum span the familiar Fock space of the
Minkowski-spacetime field.  Thus mode choice~(3.3) may be distinguished
as providing the physical vacuum/particle definition~[6] for the
field~$\varphi$ in Milne spacetime.
\vskip\baselineskip
\centerline{\bf IV.~~THE ``NATURAL'' VACUUM IN RADIATION-DOMINATED}\nobreak
\centerline{\bf FRW SPACETIME}\par\nobreak
Field mode functions for the radiation-dominated FRW geometry can be
found as functions of conformal time~$\eta$.  With
$u_\kappa(\eta)=v_\kappa(\eta)/a_R(\eta)$, Eq.~(2.6) takes the form
$${d^2v_\kappa\over d\eta^2}-[(\half\mu^2 a_*^2-\kappa^2)-
\half\mu^2 a_*^2\cosh(2\eta)]v_\kappa=0\ .\eqno(4.1)$$
This is the modified form of Mathieu's equation.  Its solutions can be
expressed in many different ways; for those which approach the Milne-spacetime
functions~$H_{i\kappa}^{(2)}(\mu t)$ the form
$$v_\kappa(\eta)=\sum_{n=-\infty}^{+\infty}c_n\,
H_{n+\nu}^{(2)}(\mu a_*\cosh\eta)\eqno(4.2)$$
is most appropriate.  In consequence of Eq.~(4.1), the coefficients~$c_n$
must satisfy the system of equations
$$\fourth\mu^2a_*^2(c_{n+2}+c_{n-2})+[\half\mu^2a_*^2-\kappa^2-(n+\nu)^2]c_n
=0\eqno(4.3)$$
for all $n$, with overall normalization determined by Eq.~(2.7).  The
necessary and sufficient condition for system~(4.3) to have a solution is
that the infinite determinant
$$\eqalignno{\Delta(\nu)&\equiv\left|\matrix{
\ldots&\ldots&\ldots&\ldots&\ldots&\ldots&\ldots\cr
\ldots&0&\qentry3&0&0&0&\ldots\cr
\ldots&1&0&\qentry2&0&0&\ldots\cr
\ldots&0&1&0&\qentry1&0&\ldots\cr
\ldots&\zentry&0&1&0&\zentry&\ldots\cr
\ldots&0&\qentry{-1}&0&1&0&\ldots\cr
\ldots&0&0&\qentry{-2}&0&1&\ldots\cr
\ldots&0&0&0&\qentry{-3}&0&\ldots\cr
\ldots&\ldots&\ldots&\ldots&\ldots&\ldots&\ldots\cr}\right|\ ,&\cr
&&(4.4)\cr}$$
with $q\equiv\fourth\mu^2a_*^2$ and $A\equiv\half\mu^2a_*^2-\kappa^2$, equal
zero.  Standard methods exist for the treatment of such determinants~[8].
The condition $\Delta(\nu)=0$ determines the parameter~$\nu$, viz.,
$$\nu={1\over\pi}\arcsin\left[\sqrt{\Delta(0)}\sin(\pi\sqrt{A})\right]
\ .\eqno(4.5)$$
The coefficients~$c_n$ can then be found, establishing the solution~(4.2).

Every term in series~(4.2) behaves asymptotically as
$$(\mu a_*\cosh\eta)^{-1/2}e^{-i\mu a_*\cosh\eta}\sim(\mu t)^{-1/2}e^{-i\mu t}
\eqno(4.6)$$
for $\eta\gg1$.  Hence the resulting solution matches the physical
positive-frequency mode function for the Milne-spacetime field determined
above, in the late-time limit.  The vacuum state and Fock space based on
this mode choice will exhibit the same regularity (lack of extraneous
singularities~[6]), analyticity, and other properties as standard
Minkowski-spacetime field theory in that limit---as indeed does quantum
field theory in the actual Universe at the present time.  In this sense,
then, this choice may be termed the ``natural'' vacuum/particle definition
in open, radiation-dominated FRW spacetime.
\vskip\baselineskip
\centerline{\bf V.~~CONCLUSIONS}
This example illustrates how a possible ``natural'' choice of
positive-frequency normal modes, vacuum, and particle states for a
quantized field can be made, specifically in an {\it open\/} or hyperbolic
FRW spacetime geometry.  That choice is based on correspondence with
standard Minkowski-spacetime theory at late times, in accord with all
of experimental particle physics.

Conversely, hyperbolic or open cosmological models can be viewed
as ``natural,'' in that they admit such a definition of vacuum and
particle states.  The construction illustrated here can be applied to
any hyperbolic FRW spacetime without cosmological constant, i.e., any
spacetime which approaches the Milne geometry at late times.  Of course
this includes the matter-dominated model to which the present Universe
more closely corresponds, though the forms of Eq.~(2.6) and its solutions
are then more unwieldy and less familiar than in the radiation-dominated
example.

Current evidence also admits the possibility that the cosmological constant
may be nonzero~[2,3].  If it is positive, then the Universe will ultimately
approach a de Sitter geometry, as its expansion suppresses all other density
contributions.  In that case a similar ``natural'' vacuum can also be chosen,
on physical grounds akin to those in the zero-cosmological-constant case:
The ``natural'' vacuum is that which approaches the Euclidean
(Chernikov-Tagirov~[9] or Bunch-Davies~[10]) vacuum in de Sitter space
at late times.  An explicit example would be more involved that that shown
above and will not be attempted here.

Though identified by their behavior at late times, the vacuum and particle
states chosen as described here can serve to describe the physical state of
the field at all times.  The mode functions used are exact, not approximate,
solutions of the field equation---equivalently, the vacuum/particle states
are exact solutions of the functional Schr\"odinger equation for the field.
Consequently, the actual state of the field is the same superposition of
these basis states at early as at late times (though physical properties
of the states may be very different).  This choice of vacuum state could
even serve as the ``preferred natural, geometrical vacuum state'' in the
normal-ordering program described by Brown and Ottewill~[11].

The dependence of this construction on the particular behavior of the
spacetime geometry suggests the possibility that the formulation of
physics might be more intimately related to the actual structure of
the Universe than is often supposed.  It may be noted, however, that
no ``teleology'' is involved in the notion that the vacuum and particle
states---hence, the state space of the field---are determined at all times
by particular late-time behavior.  Since both the spacetime geometry and
the mode functions of the field are solutions of second-order differential
equations, the late-time behavior of both is implicit in their initial
conditions.\par
\hrule
\item{[1]}C.~W.~Misner, K.~S.~Thorne, and J.~A.~Wheeler, {\it Gravitation\/}
(Freeman, San Francisco, CA, 1973), p.~704.
\item{[2]}P.~M.~Garnavich {\it et al,\/} Astrophys.~J.~{\bf 493,} L53 (1998).
\item{[3]}S.~Perlmutter {\it et al,\/} Nature (London) {\bf 391,} 51 (1998). 
\item{[4]}E.~A.~Milne, Nature (London) {\bf 130,} 9 (1932).
\item{[5]}Ref.~1, pp.~743--744
\item{[6]}S.~Winters-Hilt, I.~H.~Redmount, and L.~Parker, Phys.~Rev.~D,
submitted.
\item{[7]}Ref.~1, frontispiece.
\item{[8]}E.~T.~Whittaker and G.~N.~Watson, {\it A Course of Modern
Analysis\/} (Cambridge University Press, Cambridge, England, 1927),
pp.~36--37 and 413--417.
\item{[9]}N.~A.~Chernikov and E.~A.~Tagirov, Ann.~Inst.~Henri Poincar\'e
{\bf 9A,} 109 (1968).
\item{[10]}T.~S.~Bunch and P.~C.~W.~Davies, Proc.~R.~Soc.~London {\bf A360,}
117 (1978). 
\item{[11]}M.~R.~Brown and A.~C.~Ottewill, Proc.~R.~Soc.~London {\bf A389,}
379 (1983).
\vfill\eject\end